\documentclass[12pt]{article}
\usepackage{epsf,axodraw}
\usepackage[cp1251]{inputenc}

\begin{document}

\title{{\bf Nonlinear Regge Trajectories in Theory and Practice}}
\author{V.A. Petrov (in collaboration with A.A. Godizov)\\
{\small {\it Institute for High Energy Physics, Protvino, Russia}}\\
\\
{\footnotesize {\bf Abstract.} The problems related to nonlinear 
behavior of Regge trajectories (RT) and 
their renormalization}\\ 
{\footnotesize group invariance are discussed.}
}
\date{}
\twocolumn[\maketitle]


\section*{Spectroscopy}

{\footnotesize Since the very first papers on Regge analyses of high-energy 
behavior of 
the hadron scattering amplitudes it was a habit to use an extrapolation 
of the strikingly linear behavior of Regge trajectories at positive 
invariant mass squared, $t$, to the scattering region, 
where $t\le 0$. 
\begin{figure}[h]
\epsfxsize=7cm\epsfysize=7cm\epsffile{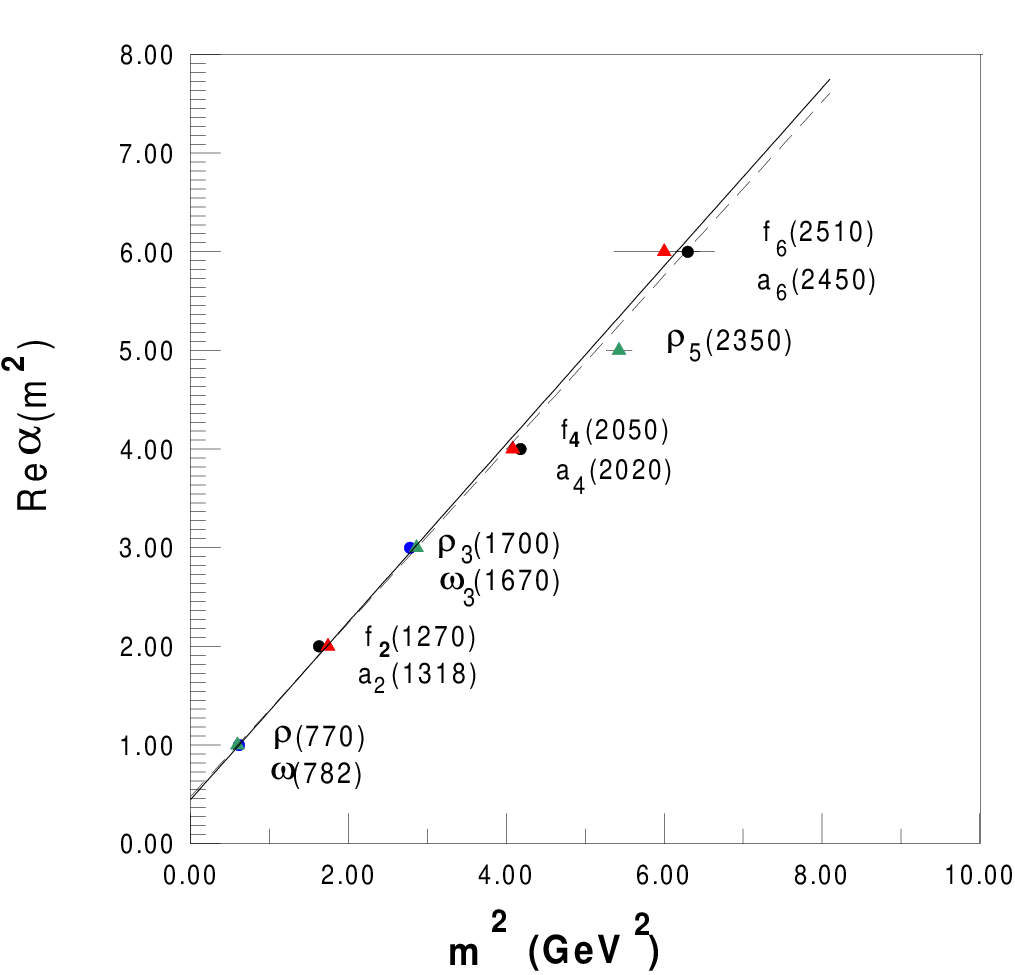}
\caption{The data on hadron spectroscopy.}
\label{spectr}
\end{figure}
Afterwards the famous Veneziano model gave an argument in favour of 
the linear and exchange degenerated trajectories, at least as a good 
first approximation, probably subject to minor nonlinear modifications. 
Fig. \ref{spectr} illustrates the Chew-Frautschi plot for $\rho$-$\omega$ 
meson family, which seemingly confirms such an idea. However, as was 
stressed in Ref. \cite{martynov}, the models implying both linearity and 
exchange degeneracy lead to an awfully bad $\chi^2/d.o.f.\ge 100$. 
Nonetheless, one can weaken the demand on RT giving up, say, exchange 
degeneracy. Does it mean that (in case if the timelike ($t>0$) region 
is described well with approximately linear trajectories) we have 
to expect the same in the scattering ($t\le 0$) region?
}

\section*{Theory}

{\footnotesize Prior to search for an apriory answer it is interesting 
to try to see what simplest field theoretic models, like $\phi^3$- or  
$\phi^4$-theories yield for the RT \cite{sawyer}.
Fig. \ref{phi} exhibits a typical picture of the RT $t$-dependence 
which can be only approximately linear in some interval of positive $t$, 
while at negative $t$ (and large enough positive $t$) flattens 
and tends to some constant, depending on the model. 
\begin{figure}[h]
\epsfxsize=9cm\epsfysize=2.5cm\epsffile{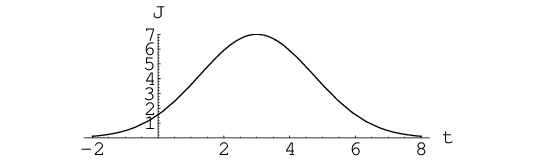}
\caption{The typical dependence of RT in simple field models.}
\label{phi}
\end{figure}

Certainly, $\phi^3$ and $\phi^4$ models are not very realistic, and 
afterwards some phenomenological nonlinear models for RT were suggested. 
We mention only three options (of many). In Ref. \cite{prokudin} the 
trajectory of the form 
\begin{equation}
\alpha(t)=1+\gamma(\sqrt{t_0}-\sqrt{t_0-t})
\label{prokud}
\end{equation}
was considered and applied to quite a successful description of the 
scattering data. One of the distinctive features of this model is negative 
and infinite value of $\alpha(t)$ at infinitely large and negative $t$, 
which is difficult to interpret physically. This trajectory is 
devoid of particles (this is not prohibited, though). 

Equation 
\begin{equation}
\alpha(t)=c\ln(b-at),\;\;c\sim g^2
\label{coo}
\end{equation}
represents the attempt to account for some QCD-like features, 
such as vector gluon exchange \cite{coon}. The resulting trajectory is 
negative at large ($\pm t$) and again devoid of particles on it. 

Finally, the equation 
\begin{equation}
\alpha(t)=a+bt+b\sqrt{(t_0-t)(t_0^*-t)}
\label{kearn}
\end{equation}
from Ref. \cite{kearney} gives us something different: the RT grows 
linearly at large ($+t$) and tends to a constant (-1) at large ($-t$). 
Linear behavior is normally related to the stringlike behavior of 
hadrons at higher spins. We remark that $\alpha(t)$ from Eq. 
(\ref{kearn}) has two complex conjugated branch points, $t_0$ and $t^*_0$.
}

\section*{QCD}

{\footnotesize If we turn to QCD -- supposedly the ultimate theory of strong 
interactions -- then we find that in the region of applicability of 
perturbative methods, i.e. large ($-t$), the pomeron trajectory tends to +1 at 
large ($-t$) \cite{kirlip}, while the secondary (meson) trajectory tends to 0 
in the same direction \cite{kwiecinski}: 

\begin{center} 
\begin{picture}(500,40)
\Line(0,40)(50,40)
\Line(0,0)(50,0)
\Photon(25,0)(25,40){3}{15}
\Vertex(25,40){2}
\Vertex(25,0){2}
\Line(75,40)(125,40)
\Line(75,0)(125,0)
\Photon(90,0)(90,40){3}{5}
\Photon(115,0)(115,40){3}{5}
\Vertex(90,40){2}
\Vertex(90,0){2}
\Vertex(115,40){2}
\Vertex(115,0){2}

\Line(150,40)(200,40)
\Line(150,0)(200,0)
\Photon(165,0)(165,40){3}{5}
\Photon(190,0)(190,40){3}{5}
\Vertex(165,40){2}
\Vertex(165,0){2}
\Vertex(190,40){2}
\Vertex(190,0){2}
\Photon(165,20)(190,20){3}{3}
\Vertex(165,20){2}
\Vertex(190,20){2}


\Text(10,15)[b]{{\Large P}}
\Text(67,17)[b]{{\Large =}}
\Text(138,15)[b]{{\Large +}}
\Text(213,15)[b]{{\Large $+...$}}

\end{picture}

\vskip -0.7cm
\begin{equation}
\alpha_P(t)\;\;\;\;\;\;\;\;\;\to \;\;\;\;\;\;\;\;\;1\;\;\;\;\;\;\;\;\;+
\;\;\;\;\;\;\;\;
O(g^2(t))\;\;\;\;\;\;\;\;\;\;\;\;\;
\label{lipa}
\end{equation}
\end{center}


\begin{center} 
\begin{picture}(500,40)
\Line(0,40)(50,40)
\Line(0,0)(50,0)
\Photon(25,0)(25,40){3}{15}
\Vertex(25,40){2}
\Vertex(25,0){2}
\Line(75,40)(125,40)
\Line(75,0)(125,0)
\ArrowLine(90,0)(90,40)
\ArrowLine(115,40)(115,0)
\Vertex(90,40){2}
\Vertex(90,0){2}
\Vertex(115,40){2}
\Vertex(115,0){2}

\Line(150,40)(200,40)
\Line(150,0)(200,0)
\ArrowLine(165,0)(165,20)
\ArrowLine(190,40)(190,20)
\ArrowLine(165,20)(165,40)
\ArrowLine(190,20)(190,0)
\Vertex(165,40){2}
\Vertex(165,0){2}
\Vertex(190,40){2}
\Vertex(190,0){2}
\Photon(165,20)(190,20){3}{3}
\Vertex(165,20){2}
\Vertex(190,20){2}

\Text(10,15)[b]{{\Large R}}
\Text(67,17)[b]{{\Large =}}
\Text(138,15)[b]{{\Large +}}
\Text(213,15)[b]{{\Large $+...$}}

\end{picture}

\vskip -0.6cm
\begin{equation}
\;\;\;\alpha_R(t)\;\;\;\;\;\;\;\;\;\;\to \;\;\;\;\;\;\;0\;\;\;\;\;\;\;+
\;\;\;\;\;\;\;\;
O(g(t))\;\;\;\;\;\;\;\;\;\;\;\;\;
\label{kwie}
\end{equation}
\end{center}

Such an asymptotic corresponds to exchanges of non-interacting gluons 
and quarks. This seems to be quite natural because at small distances 
(large $-t$) we expect the effects of asymptotic freedom. 
However, the approach to these asymptotic values appears to be quite 
nonperturbative (terms of order $g^{10/3}$ for the pomeron trajectory 
and $g^{5/3}$ for the meson one). 
Both the results were obtained via a Bethe-Salpeter equation 
with the kernel corresponding to the lowest order in QCD coupling. 

What do we get at small $t$? Leading singularity $\alpha_{gg}(0)$ 
was calculated in Ref. \cite{camici} in two first orders in the QCD 
coupling $g^2$, 
\begin{equation}
\alpha_P(0)=1+\frac{3\ln 2}{\pi^2}g^2\left(1-\frac{5}{\pi^2}g^2\right),
\label{cami}
\end{equation}
and does not contain, in contrast to the case of large ($-t$), 
nonanalytic terms in $g$. However, Eq. (\ref{cami}) 
exhibits a function which depends both on implicit scale 
dependence of the QCD coupling and the renormalization scheme. 
}

\section*{Renormalization group invariance}

{\footnotesize The last circumstance seems quite worrying because the 
Regge trajectories are generally related to physical amplitudes and 
spins and masses of observed particles. Hence they cannot depend on any 
arbitrary choice of renormalization scale and scheme. 
From the RG invariance it follows that in case of massless QCD (which 
fairly admits the existence of massive hadrons due to ``dimensional 
transmutation'') RT have a general form 
\begin{equation}
\alpha(t)=\Phi\left(\frac{t}{\mu^2}e^{K(g^2)}\right)
\label{genren}
\end{equation}
with the renormalization scale $\mu$ and QCD coupling $g$. 
$\Phi$ is some function that has to be defined by the dynamics 
and $dK/dg^2=\beta(g^2)$ is the usual QCD $\beta$-function: 
$\beta(g^2)=-\beta_0g^4-\beta_1g^6-...$

One can immediately derive from Eq. (\ref{genren}) the following 
important consequences. 
\begin{enumerate}
\item $\alpha(0)=\Phi(0)$, i.e. the intercept does not depend on 
$g^2$ at all.
\item If $\alpha(t)$ is analyic at $t=0$ then 
$\alpha(t)=\alpha(0)+\alpha'(0)t+...$ where 
\begin{equation}
\left.\alpha'(0)\right|_{g^2\to 0}
\sim\left(\frac{1}{g^2}\right)^{\frac{\beta_1}{\beta_0^2}}
e^{\frac{1}{\beta_0g^2}}.
\label{nakl}
\end{equation}
\end{enumerate}
Eq. (\ref{nakl}) shows that the slope of the RT is highly nonperturbative. 
Let us note that the independence of the intercept of $g^2$ is 
self-consistent in the sense that if one takes $\mu^2=-t$, then 
$\alpha(t)$ becomes the function of the running coupling at ($-t$):
$$
\alpha(t)=\Phi\left(-e^{K(g^2(-t))}\right).
$$
From the definition of $K$ it follows that 
$$
K(g^2(t))-K(g^2)=\ln\left(\frac{-t}{\mu^2}\right)\to-\infty\;\;(t\to 0),
$$
so we obtain again $\Phi(0)$. In the case of finite $g^2(0)$ \cite{solov}
it has to be an IR fixed point: $\beta[g^2(0)]=0$. 

So one can conclude that, first, QCD dictates essentially nonlinear RT, 
and nonlinearity, as will be seen later, is not negligible at quite low $t$; 
second, the very RT are not analytic in coupling constants which is a direct 
consequence of the RG invariance. 
}

\section*{Practice}

\begin{figure}[h]
\vskip -1cm
\hskip 1.5cm
\epsfxsize=5cm\epsfysize=5cm\epsffile{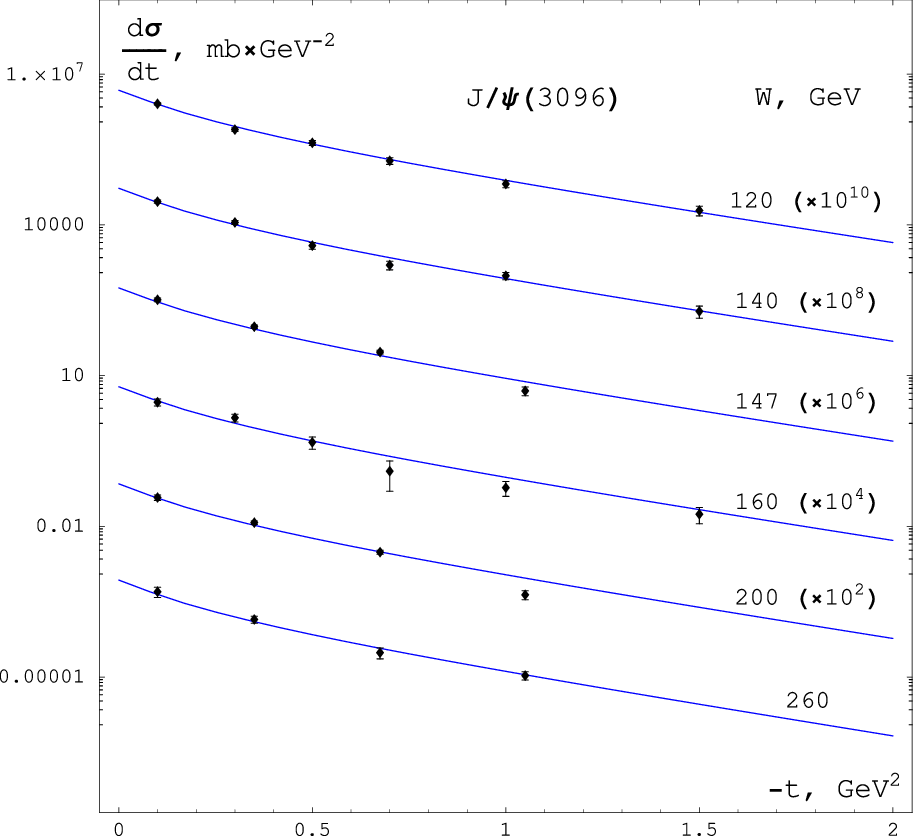}
\caption{Angular distributions for the exclusive 
$J/\psi(3096)$ photoproduction.}
\label{jpsi}
\end{figure}

{\footnotesize 

Now we turn to practical implementations of these considerations 
\cite{godizov}. 
\begin{figure}[h]
\hskip 1.5cm
\epsfxsize=5cm\epsfysize=5cm\epsffile{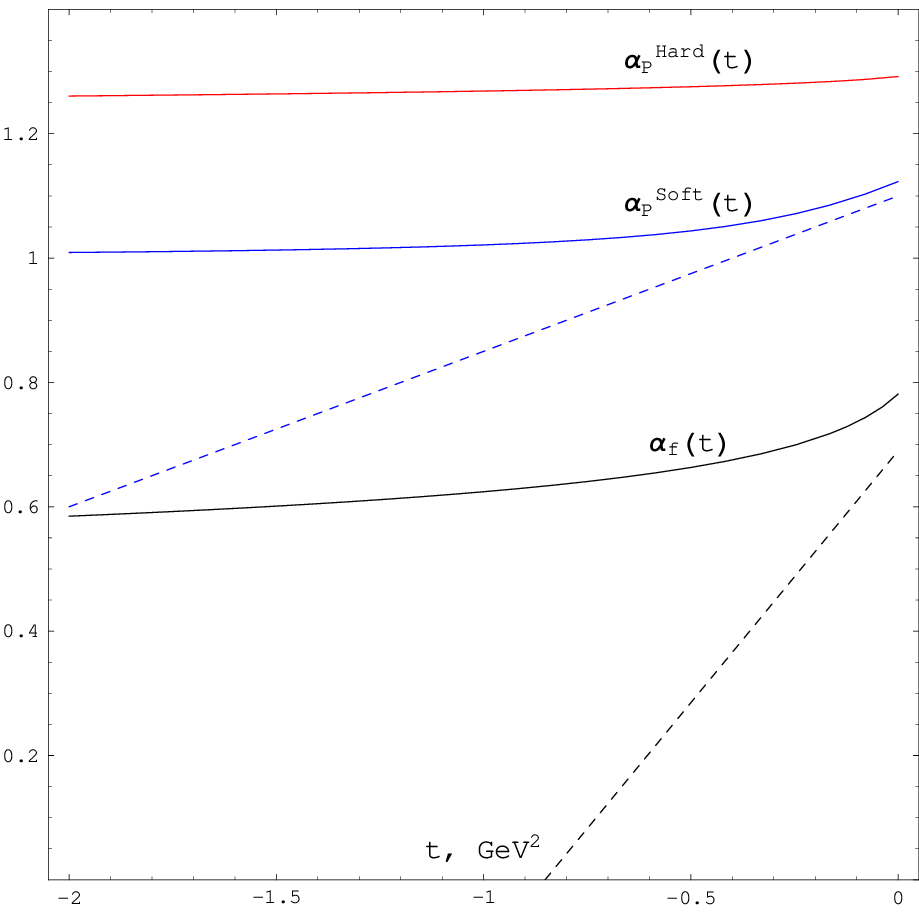}
\caption{Nonlinear leading Regge trajectories.}
\label{tra}
\end{figure}
Fig. \ref{tra} shows the $t$-dependence of the hard and soft pomeron 
trajectories, and that of the $C$-even secondary $f_2$-trajectory 
which takes into account QCD results. The dashed lines show 
the linear extrapolations from the positive $t$ region.

Nonlinear trajectories with QCD-motivated behavior were used 
also for the description of the $pp$ and $\bar p p$ scattering. 
Fig. \ref{pp} shows quite a good fit. The dashed lines show what happens 
if one takes a linear approximation for the RT. 

When one deals with hard diffractive processes the soft pomeron appears 
to be unsufficient. Thereof the need in ``hard'' pomeron. Fig. \ref{jpsi} 
demonstrates the use of combined nonlinear hard and soft pomerons with 
quite a successful quality. 

\begin{figure}[h]
\vskip -0.5cm
\hskip 1.5cm
\epsfxsize=5cm\epsfysize=5cm\epsffile{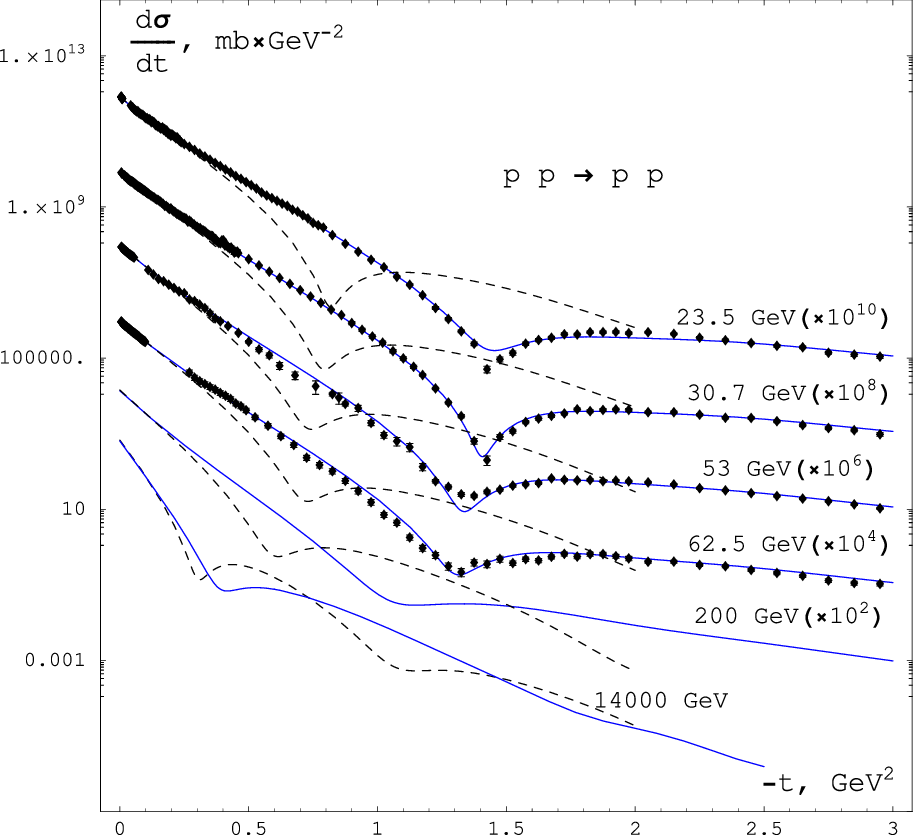}
\end{figure}
\begin{figure}[h]
\vskip -1cm
\hskip 1.5cm
\epsfxsize=5cm\epsfysize=5cm\epsffile{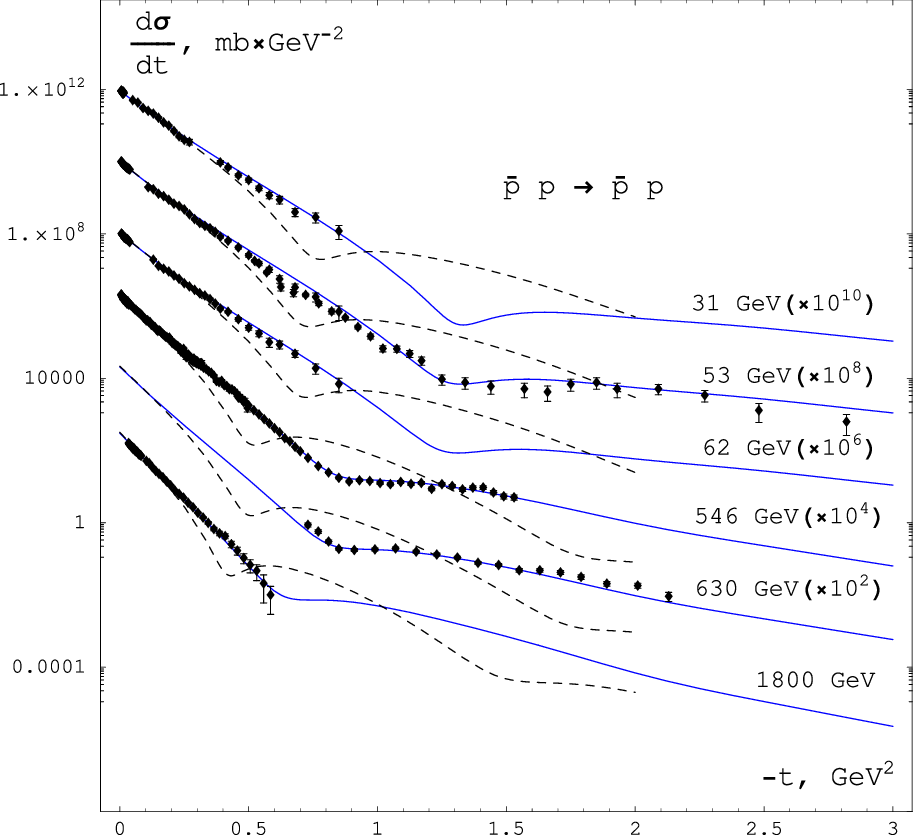}
\caption{Angular distributions for the high energy 
nucleon-nucleon scattering.}
\label{pp}
\end{figure}

}


\section*{Problems}

\begin{figure}[h]
\vskip -0.5cm
\hskip 1.5cm
\epsfxsize=5cm\epsfysize=5cm\epsffile{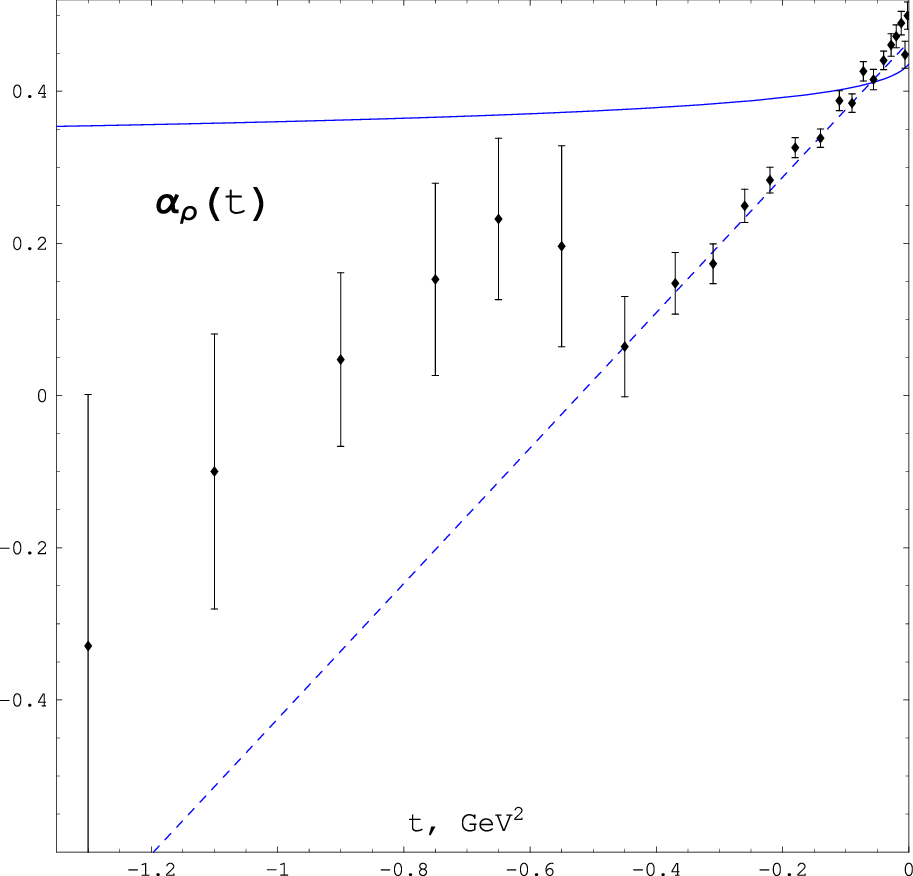}
\caption{The $\rho$-reggeon trajectory in three different approaches.}
\label{rho}
\end{figure}

{\footnotesize Certainly, there remain a lot of problems. For instance, the 
$\rho$-meson trajectory which according to QCD has to tend to zero 
at large-$t$ shows quite indefinite picture (Fig. \ref{rho}). The problem 
was addressed in Refs. \cite{brodsky,godizov2}. More exacerbated situation 
occurs with $\pi$-meson trajectory which is definitely negative already 
at $t=0$. 

If to take the QCD-motivated and physically appealing RT behavior 
depicted at Fig. \ref{pom} (parton exchanges at large ($-t$) and 
``stringy'' behavior at large $t$), then we encounter a possible 
problem with analyticity, as such trajectories have complex 
$t$-plane singularities which can be dangerous for the whole amplitude, 
regular at complex $t$ (on the first sheet). In its turn this can be 
related to possible microcausality violation \cite{samokh}, 
one of the pillars of modern physics. 
}

\newpage

\begin{figure}[h]
\vskip -0.5cm
\hskip 1.5cm
\epsfxsize=5cm\epsfysize=5cm\epsffile{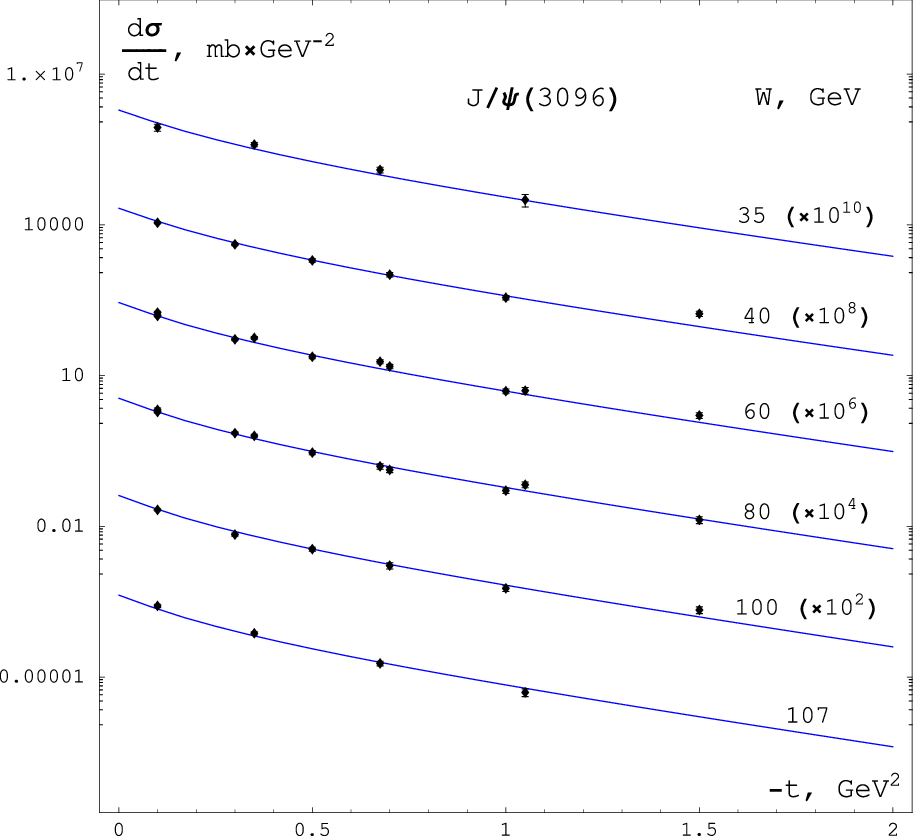}
\end{figure}

\vspace*{-1cm}

\section*{Summary}

{\footnotesize 

\begin{enumerate}

\item Analyticity of RT in $t$ seems to imply their singular behavior 
at $g^2\sim 0$;

\item QCD (partonlike) asymptotics at high $(-t)$ and ``stringy'' 
asymptotics at high $t$ imply complex singularities of RT in the $t$-plane. 
Probable clash with microcausality.

\item QCD behavior at high $(-t)$ does not match with monotony of RT in 
t for $\rho$, $\pi$,..., heavy quarkonia.

\end{enumerate}

\newpage

\begin{figure}[h]
\hskip 2cm
\epsfxsize=5cm\epsfysize=3cm\epsffile{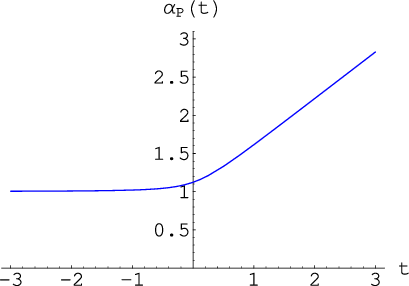}
\caption{Regge trajectory with a combined behavior.}
\label{pom}
\end{figure}

I take this opportunity to thank J. Soffer and R. Fiore, 
the organizers of this marvellous workshop, for their 
kind hospitality and essential support.

}



\end{document}